\documentclass
[twocolumn,aps,prc,amsmath,amssymb,floatfix]
{revtex4-1}
\usepackage{CJK}                     
\usepackage[dvips]{graphicx}
\usepackage{bm}                      
\usepackage{mathptmx}                
\usepackage{dcolumn}                 
\usepackage{multirow}                
\usepackage{xcolor}                  %

\usepackage{hyperref}
\hypersetup{breaklinks=true,colorlinks=true,    
linkcolor=blue,citecolor=blue,filecolor=magenta,urlcolor=black,anchorcolor=blue}

\def\be{\begin{equation}}
\def\ee{\end{equation}}
\def\bal#1\eal{\begin{align}#1\end{align}}
\def\rv{\bm{r}}
\def\eps{\varepsilon}
\def\fm3{\,\text{fm}^{-3}}
\def\gc3{\,\text{g/cm}^3}
\def\mfm{\;\text{MeV}\,\text{fm}^{-3}}

\def\kni{$K^-N$~interaction}

\begin{document}


\title{The drip lines of kaonic nuclei}


\author{J. Guo}
\affiliation{Department of Physics, East China Normal University,
Shanghai 200241, China}
\author{D. H. Chen}
\affiliation{Department of Physics, East China Normal University,
Shanghai 200241, China}
\author{Xian-Rong Zhou}\email{xrzhou@phy.ecnu.edu.cn}
\affiliation{Department of Physics, East China Normal University,
Shanghai 200241, China}
\author{H.-J. Schulze}
\affiliation{INFN Sezione di Catania, Dipartimento di Fisica,
Universit\'{a} di Catania, Via Santa Sofia 64, 95123 Catania, Italy}
\author{Q. B. Chen}
\affiliation{Department of Physics, East China Normal University,
Shanghai 200241, China}

\date{\today}

\begin{abstract}
The effects of an additional $K^-$ meson on the neutron and proton drip
lines are investigated within Skyrme-Hartree-Fock approach
combined with a Skyrme-type kaon-nucleon interaction.
While an extension of the proton drip line is observed
due to the strongly attractive $K^-p$ interaction,
contrasting effects (extension and reduction) on the neutron drip line
of Be, O, and Ne isotopes are found.
The origin of these differences is attributed to the behavior
of the highest-occupied neutron single-particle levels
near the neutron drip line.
\end{abstract}

\maketitle

\section{Introduction}
\label{intro}

Kaonic nuclei,
bound states of a negatively-charged kaon $K^-$ and a normal nucleus,
can provide valuable information concerning the
kaon-nucleon interactions at low energies.
Since Kishimoto suggested in 1999 the existence of $K^-$ nuclei
with a medium atomic number
based on the strong attraction in the $K^-N$ system \cite{Kishimoto1999PRL},
and Akaishi and Yamazaki discussed the possibility of $K^-$ nuclei
in few-body systems in 2002 \cite{Akaishi2002PRC},
$K^-$ nuclei have become one of the important problems
that attracted significant attention
from both experimental and theoretical sides
in the study of strangeness physics.

In the past fifteen years,
the properties of $K^-$ nuclei have been studied actively in several experiments.
For example,
in 2005 the FINUDA collaboration succeeded to detect a kaon-bound state $K^-pp$
through its two-body decay into a $\Lambda$ hyperon and a proton
and determined the corresponding binding energy and decay width
\cite{Agnello2005Evidence}.
Later on in 2010,
a deeply bound and compact $K^-pp$ state formed in the
$pp \rightarrow K^+ + K^-pp$
reaction was found by analyzing the data of the DISTO experiment on
the exclusive
$pp\rightarrow p\Lambda K^+$ reaction \cite{PhysRevLett.104.132502}.
Its mass and width were also obtained.
Recently, the bound state $K^-pp$ was detected by the E15 and E27 collaborations
at J-PARC.
The E15 collaboration found a bound state $K^-pp$ with a binding energy of
$47 \pm 3^{+3}_{-6}\,$MeV \cite{Sada:2016nkb}.
The E27 collaboration observed a deeply bound state $K^-pp$
produced by the reaction
$\pi^+ + n + p \rightarrow K^-pp +K^+ \rightarrow \Sigma^0 + p + K^+$
\cite{Ichikawa:2015vpa}.
As is noted, the main focus of current experiments
is on the simplest kaonic nucleus $K^-pp$;
thus searching for new kaonic nuclei in the future is necessary,
and of course,
can open up lots of new opportunities for theoretical studies.

On the theoretical side,
various models were used to study the properties of $K^-$ nuclei.
For example,
the antisymmetrized molecular dynamics model \cite{PhysRevC.70.044313},
the effective chiral Lagrangian for the kaon-baryon interaction
combined with a nonrelativistic baryon-baryon interaction model
\cite{Muto:2008fc},
the chiral SU(3) model \cite{PhysRevC.79.014003},
the chiral meson-baryon coupled-channels interaction models
\cite{PhysRevC.96.015205},
the non-relativistic Faddeev and Faddeev-Yakubovsky calculations
\cite{Shuji,Marri:2016qsf}.
For the density functional theory,
several relativistic mean field (RMF) models
explored many properties of light $K^-$~nuclei,
such as level inversion,
binding energy, decay widths,
and density distributions
\cite{PhysRevC.60.024314,PhysRevC.74.034321,PhysRevC.76.055204,
PhysRevC.79.035207,yang2014relativistic}.
Within Skyrme-Hartree-Fock (SHF) model,
we studied the potential energy surfaces, mean fields,
and the density distributions of $K^-$ nuclei
for various effective \kni\ strengths \cite{zhou2013kaonic,jin2019deformed}.

One notes that these pioneering works focussed mainly
on the kaon removal energy and width.
However, a general understanding of strange nuclear systems
requires the evaluation of the global strange nuclear chart,
with strangeness as the third dimension.
In particular, the location of the nuclear drip lines
is an important information to understand the limits of the strong force
to hold together the nucleons in a bound system.
There are many studies concerning the drip lines in normal nuclear systems,
but much less works regarding strangeness.
For example, for the $\Lambda$
\cite{PhysRevC.98.024316,
PhysRevC.78.054311,PhysRevC.84.014328,PhysRevC.89.044307,PhysRevC.90.064302,
PhysRevC.91.024327,PhysRevC.95.034309,PhysRevC.91.054306,PhysRevC.95.024323,
PhysRevC.97.034302,Umeya_2020,fang2020impurity},
$\Sigma$
\cite{HARADA2005143,PhysRevC.58.3688,PhysRevC.97.024601,RevModPhys.88.035004,
PhysRevC.98.024316},
and $\Xi$
\cite{PhysRevC.78.054316,PhysRevC.94.064319,PhysRevC.49.2472,tan2004different,
PhysRevC.98.024316,jin2020study} hypernuclei,
the potentials, single-hyperon levels, deformation, density distributions,
and binding energies have been investigated extensively.
But only for $\Lambda$ hypernuclei the neutron drip lines were investigated,
e.g., in Refs.~\cite{PhysRevC.78.054306,Samanta2008JPG,PhysRevC.92.044313}.
Definitely, the prediction of the drip lines in $K^-$ nuclei is
of importance to explore the strange nuclear chart in the future.

Therefore, the present work aims to provide the first investigation
of the effects of an additional $K^-$ on the neutron and proton drip lines
in some typical $K^-$ nuclei within the SHF approach.
We take the isotopes of the typical nuclei Be, O, and Ne as examples.
While most of Be and Ne isotopes are deformed,
the deformations of O isotopes are very small.

The paper is organized as follows.
In Sec.~\ref{s:form}, the extended SHF
approach as well as the Skyrme-type $K^-N$ force are introduced.
In Sec.~\ref{s:res}, the energies of the highest-occupied
single-particle levels, one-nucleon separation energies, binding energies, and quadrupole deformations
of $K^-$ nuclei in Be, O, and Ne isotopes are presented
together with the corresponding normal nuclei
and the available experimental data.
Finally, a summary is given in Sec.~\ref{s:end}.

\section{Theoretical framework}
\label{s:form}

The main purpose of the present work is to discuss the effect of an
additional $K^-$ meson on the neutron and proton drip lines in the SHF approach
combined with a simple density-dependent Skyrme-type \kni.
In this approach
the total energy of the nucleus can be expressed as
\cite{PhysRevC.5.626,PhysRevC.7.296,Rayet:1981uj,Rayet:1976fs,Bender:2003jk,
Stone:2006fn,PhysRevC.90.047301,zhou2013kaonic,jin2019deformed}
\be
\label{e:energy}
 E = \int d^3r\, \eps(\rv) \:,\quad
 \eps = \eps_{NN} + \eps_{KN} + \eps_C \:,
\ee
where $\eps_{NN}$ denotes the energy density of the nucleon-nucleon part,
$\eps_{KN}$ is the energy density due to the kaon-nucleon interaction,
and $\eps_C$ is the Coulomb contribution of protons and kaons.

For each single-particle (s.p.) state $\phi_q^i$ $(q=n,p,K)$,
the minimization of the total energy $E$ in Eq.~(\ref{e:energy})
implies the SHF Schr\"odinger equation
\be
 \Big[ -\nabla \cdot \frac{1}{2m^*_q(\rv)} \nabla + V_q(\rv)
 - \bm{W}_q(\rv) \cdot (\nabla\times\bm{\sigma}) \Big] \phi_q^i(\rv)
 = e_q^i \phi_q^i(\rv) \:,
\label{e:se}
\ee
with the mean fields
\bal
\label{e:Vk}
 V_K &= \frac{\partial\eps_{KN}}{\partial\rho_K} - V_C \:,
\\
\label{e:Vq}
 V_q &= V^\text{SHF}_q + V^{(K)}_q \:,\
 V_q^{(K)} = \frac{\partial\eps_{KN}}{\partial\rho_q} ,\ (q=n,p) \:,
\eal
where
$V_C$ denotes the Coulomb field,
$V^\text{SHF}_q$ the standard nucleonic Skyrme mean field,
$\bm{W}_q$ the nucleonic spin-orbit mean field,
and $V_q^{(K)}$ the change of the nucleonic mean fields by the \kni.

For the nucleonic part, we use the Skyrme force SLy4.
For the kaonic energy-density contribution,
a simple linear energy density functional
is assumed as in Ref.~\cite{jin2019deformed},
\be
 \eps_{KN} = -a_0 \rho_K [(1+x_0)\rho_p + (1-x_0)\rho_n] \:,
\ee
where $a_0$ and $x_0$ are the \kni\ strength parameters.
Under this assumption,
the mean fields in Eqs.~(\ref{e:Vk})
and (\ref{e:Vq}) are~\cite{jin2019deformed}
\bal
\label{e:a0a}
 V_K &= -a_0 [(1+x_0)\rho_p + (1-x_0)\rho_n] - V_C \:,
\\
\label{e:a0b}
 V_{p,n}^{(K)} &= -a_0(1 \pm x_0)\rho_K \:.
\eal
In the following calculations,
we use a $(p,n)$-symmetric \kni,
i.e., $x_0=0$ and $a_0=500\mfm$,
which were justified as reasonable values in \cite{jin2019deformed}.

The pairing interaction of the nucleonic part employs a
density-dependent $\delta$ pairing force~\cite{Tajima:1993svu},
\be
 V_q(\rv_1,\rv_2) =
 -V_0 \left[1-\frac{\rho_N((\rv_1+\rv_2)/2)}{\rho_0}\right]
 \delta(\rv_1-\rv_2) \:,
\ee
with the pairing strength $V_0=410\mfm$
for both neutrons and protons
\cite{PhysRevC.83.014301,PhysRevC.70.054316,PhysRevC.72.054311,
fang2020impurity,bender2000pairing}
and a saturation density $\rho_0=0.16\fm3$.
A smooth energy cutoff is also
included in the BCS calculation \cite{bender2000pairing}.

We assume axially-symmetric mean fields and the properties of axially-deformed
nuclei are studied in cylindrical coordinates.
The solutions of the SHF equation
are expanded within a axially-deformed harmonic-oscillator basis
\cite{PhysRevC.7.296,Bender:2003jk,Stone:2006fn}.

\section{Results and discussions}
\label{s:res}

In order to study the effect of an additional $K^-$ meson on the nucleon
drip line in detail,
we will examine the one-nucleon separation energies
\bal
 S_n &\equiv E[^AZ] - E[^{A-1}Z]\:,
\\
 S_p &\equiv E[^AZ] - E[^{A-1}(Z-1)]  \:,
\eal
which determine the drip lines and we will compare the results obtained for
normal and kaonic nuclei.
Also, we focus on the highest-occupied (valence) nucleon s.p.~levels,
which would become very weakly bound close to the drip lines.
If the s.p.~energy of the highest-occupied nucleon levels is
still negative in the minimum of the total energy,
the nucleus is supposed to exist \cite{PhysRevC.78.054306}.

\begin{figure}[t]
	\centerline{\includegraphics[width=83mm]{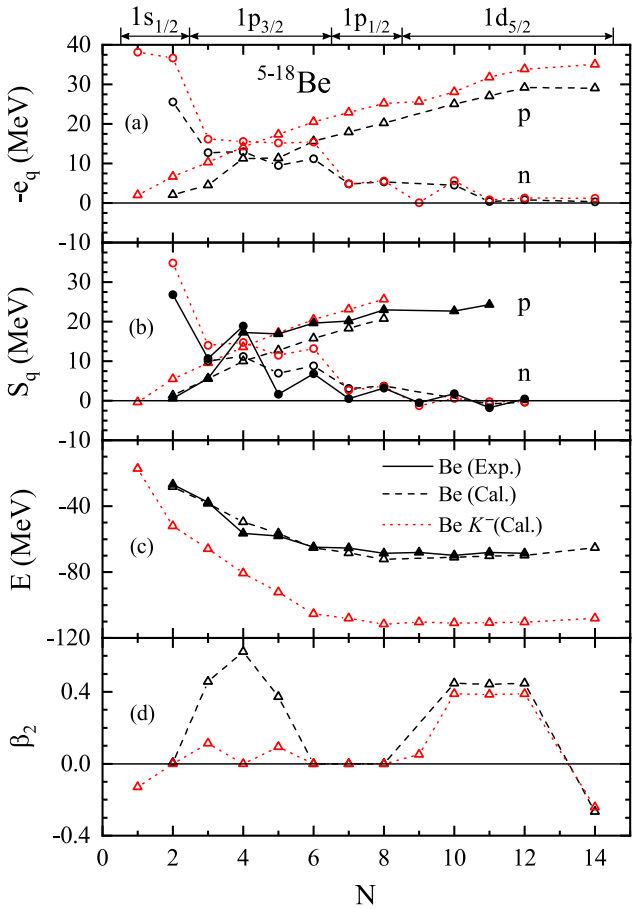}}
	\vspace{-2mm}
	\caption{
		(a) Energies of the highest-occupied nucleon s.p.~levels
		$-e_q$ ($q=n,p$)
		(the neutron level is indicated on the top;
		the proton level is $1p_{3/2}$),
		(b) one-nucleon separation energies $S_q$,
		(c) binding energies $E$,
		(d) quadrupole deformations $\beta_2$
		of Be isotopes (dashed lines)
		and their corresponding $K^-$~nuclei (dotted lines),
		obtained with $a_0=500\mfm$ and $x_0=0$ in Eqs.~(\ref{e:a0a},\ref{e:a0b}).
		The experimental $S_q$ and $E$ values of normal nuclei obtained
		from Ref.~\cite{nndc} are also included for comparison (solid lines).
	}
	\label{f:be}
\end{figure}

In Fig.~\ref{f:be},
the energies of the highest-occupied nucleon s.p.~levels $-e_q$ (a),
the one-nucleon separation energies $S_q$ (b),
the binding energies $E$ (c),
and the quadrupole deformations $\beta_2$ (d)
of Be isotopes and their corresponding $K^-$~nuclei
are presented in comparison with the available experimental results
from Ref.~\cite{nndc}.
As indicated by the theoretical $e_q$ and $S_q$ values,
the nuclei with $N=2$--$8$ neutrons and their corresponding $K^-$~nuclei exist.
However, $^{13,17,19}$Be ($N=9,13,15$) are unbound systems due to pair breaking.
Therefore, the neutron drip line locates at $^{12}$Be ($N=8$).
The proton drip line reaches at $^6$Be ($N=2$).
Experimentally, the nuclei from $^6$Be ($N=2$) to $^{16}$Be ($N=12$)
have been observed \cite{nndc},
but the experimental values of $S_n$ for $^{12}$Be ($N=8$)
and $S_p$ for $^7$B ($N=2$) are negative.
Therefore, the proton and neutron drip lines locate at $^6$Be ($N=2$)
and $^{12}$Be ($N=8$), respectively.

\begin{figure}[t]
\centerline{\includegraphics[width=83mm]{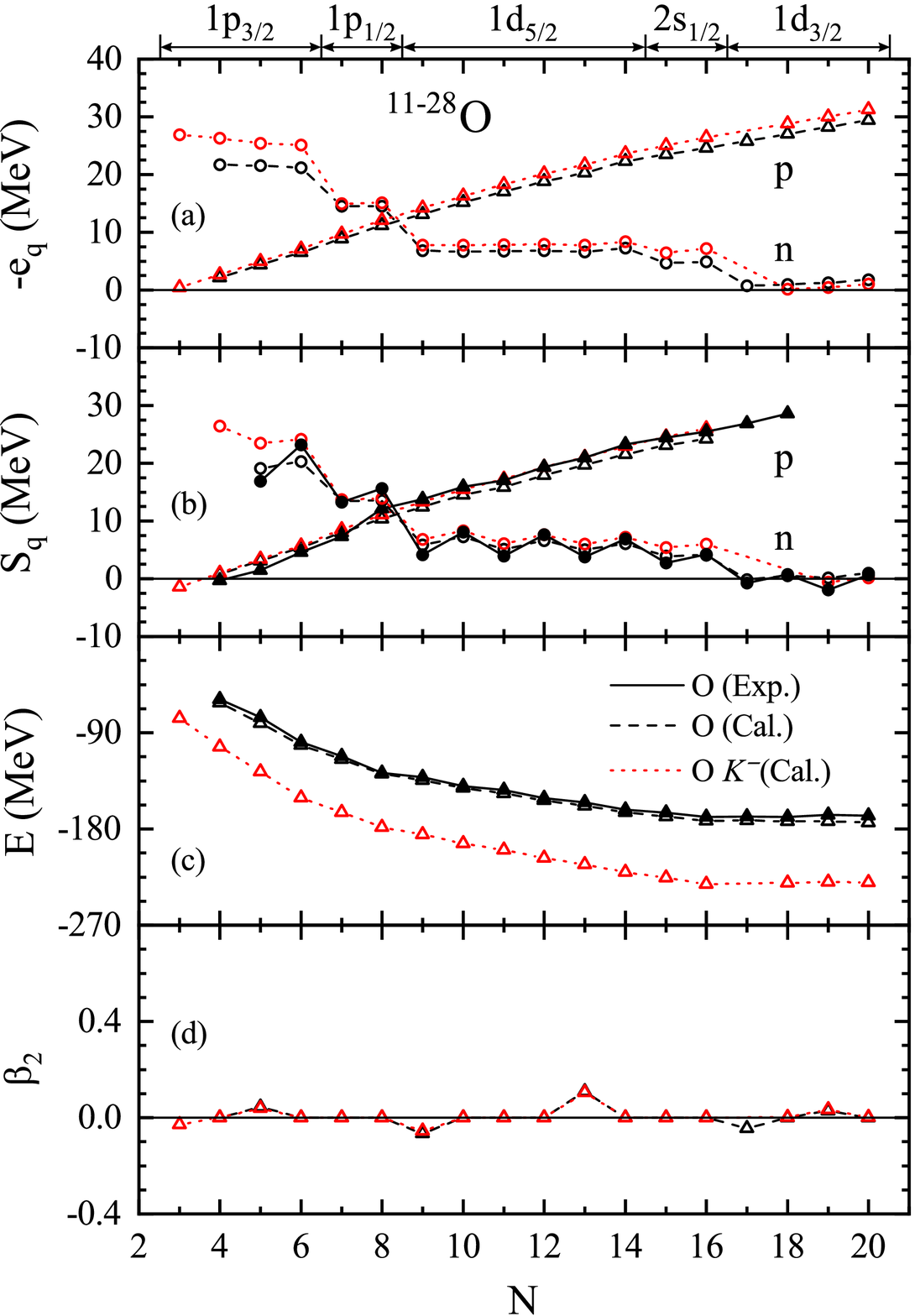}}
\vspace{-2mm}
\caption{
Same as Fig.~\ref{f:be}, but for O isotopes.
}
\label{f:o}
\end{figure}

\begin{figure}[t]
\centerline{\includegraphics[width=83mm]{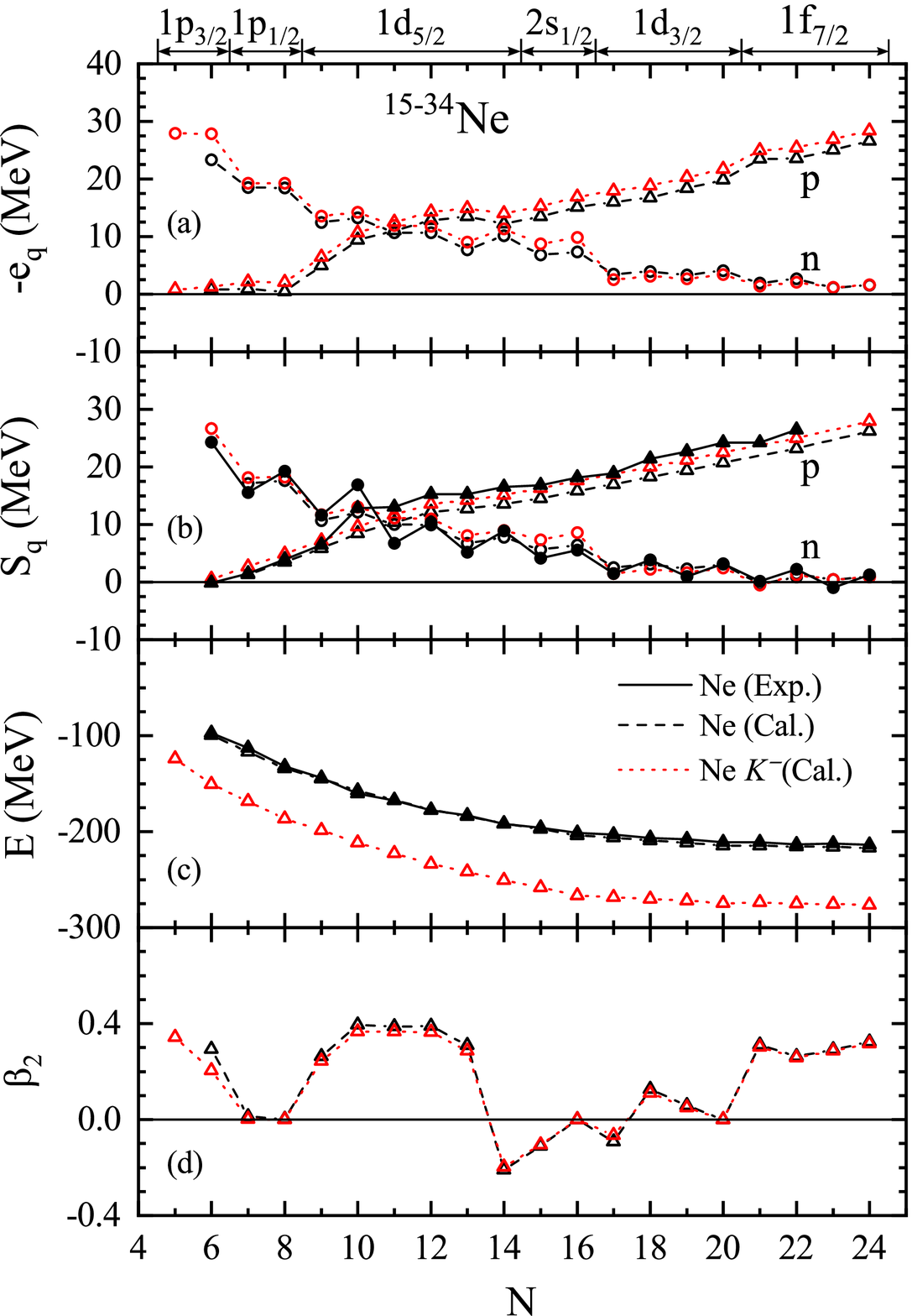}}
\vspace{-2mm}
\caption{
Same as Fig.~\ref{f:be}, but for Ne isotopes.
}
\label{f:Ne}
\end{figure}

The additional $K^-$ meson shifts down clearly the energies of the
highest-occupied nucleon s.p.~levels $e_q$
(most notable for those in strongly bound inner orbits)
and thus increases the total binding energies.
The shift for the proton levels is larger than that for the neutron levels,
even for weakly bound states,
which is due to the increased Coulomb field.
The decrease of quadrupole deformations,
as shown in Fig.~\ref{f:be}(d),
is due to the attractive \kni~\cite{jin2019deformed}.
With an additional $K^-$ meson,
$^{17}_{K^-}$Be ($N=13$) and $^{19}_{K^-}$Be ($N=15$)
remain unbound,
while $^{13}_{K^-}$Be ($N=9$) becomes marginally bound,
but its $S_n$ is still negative.
The neutron drip line point $N=8$ for Be isotopes is thus not shifted
by an additional $K^-$ meson.
Moreover, due to the major impact on the proton levels,
a bound nucleus $^{\ \ 5}_{K^-}$Be ($N=1$) is found,
but its one-proton separation energy $S_p$ remains negative.
An extension effect of an additional $K^-$ meson on the proton drip line
is thus found,
which can be attributed to the attraction of the added Coulomb
interaction between kaon and proton.
A similar phenomenon was also found by additional $\Lambda$ hyperons
due to the attractive $\Lambda N$ interaction \cite{PhysRevC.78.054306}.

Fig.~\ref{f:o} shows the preceding results for $^{11-28}$O ($N=3-20$)
as well as the corresponding $K^-$~nuclei.
In this case all isotopes are subsphaeroidal
as illustrated in panel (d).
The theoretical results of the total energies are in quite good agreement
with the experimental values.
All isotopes from $^{12}$O ($N=4$) to $^{28}$O ($N=20$) exist
with negative $e_n$ and $e_p$.
The separation energies $S_n$ and $S_p$ of all nuclei with $N=5-16$
are positive,
whereas $S_n$ becomes negative at $N=17$ theoretically
and experimentally \cite{nndc},
and thus the neutron drip line locates at $^{24}$O ($N=16$).
The theoretical proton drip line is reached at $N=4$ with
$^{13}$F being unbound,
whereas experimentally it lies at $N=5$.
The experimental one-proton separation energy $S_p$ of $^{14}$F is negative.

In contrast to the results of Be isotopes in Fig.~\ref{f:be},
the additional $K^-$ meson increases slightly the energies of the
highest-occupied neutron s.p.~levels of
$^{11-24}_{\hskip1em K^-}$O ($N=3$--$16$),
but decreases those of $^{25-28}_{\hskip1em K^-}$O
(valence neutron levels $1d_{3/2}$).
As a consequence, $^{25}_{K^-}$O becomes unbound and the
neutron removal energies of $^{27-28}_{\hskip1em K^-}$O are reduced.
Thus, a reducing effect of the $K^-$ on the neutron drip line for O isotopes
is found.
This is an interesting result and will be analyzed in detail in the following.
On the other hand, a bound $^{11}_{K^-}$O
and increasing one-proton separation energies of $^{12-24}_{\hskip1em K^-}$O
are found.
Thus, an extension effect of $K^-$ meson on the proton drip line is obvious.

\begin{figure}[t]
\centerline{\includegraphics[width=84mm]{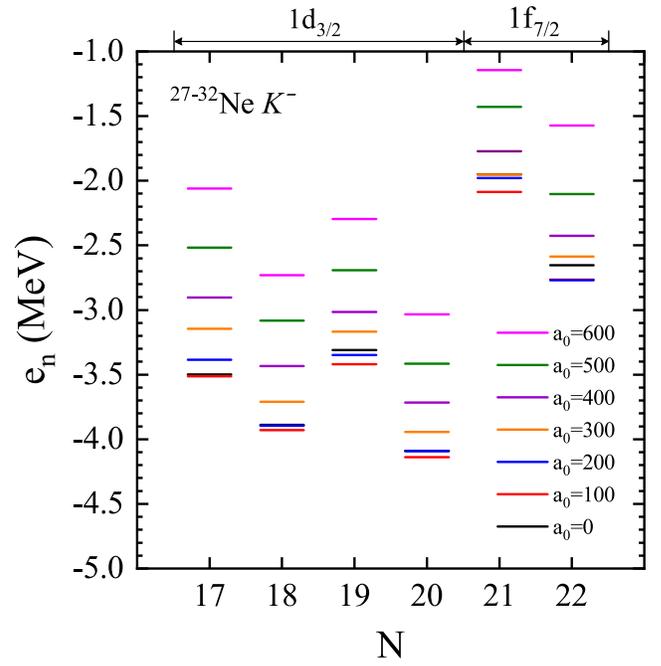}}
\vspace{-2mm}
\caption{
Energies of the highest-occupied neutron s.p.~levels
of $^{27-32}_{\hskip0.99em K^-}$Ne
and their corresponding normal nuclei $(a_0=0)$
obtained with different \kni\ strengths~$a_0$
(in units of $\mfm$).
}
\label{f:en}
\end{figure}

The same phenomenon occurs for Ne isotopes,
as observed in Fig.~\ref{f:Ne}.
All isotopes $^{16-34}$Ne ($N=6-24$) exist, as experimentally~\cite{nndc}.
$^{35,36}$Ne are unbound systems.
For $^{17-30}$Ne ($N=7$--$20$),
$S_n$ and $S_p$ are positive both experimentally and theoretically,
while $S_n$ of $^{31}$Ne ($N=21$) is negative theoretically
but marginally positive experimentally.
Thus, the neutron drip line locates at $^{30}$Ne ($N=20$) theoretically
and at $^{32}$Ne ($N=22$) experimentally.
The proton drip line is not reached
for nuclei with $N>6$ experimentally and theoretically.
A weakly bound $^{16}$Ne ($Z=10,N=6$)
and an unbound $^{15}$F ($Z=9,N=6$)
are found on the theoretical side,
while the experimental $S_p$ of $^{16}$Ne is slightly negative \cite{nndc}.

However, $^{16}_{K^-}$Ne with positive $S_p$ and $^{15}_{K^-}$F with
negative $S_p$ are bound nuclei and thus the additional $K^-$ meson
firmly establishes the proton drip line at $^{14}$O for $N=7$ nuclei,
and in fact also $^{15}_{K^-}$Ne becomes bound.
All kaonic nuclei $^{15-34}_{\hskip1em K^-}$Ne ($N=5$--$24$) exist.
Thus,
the additional $K^-$ meson does not affect the neutron drip line of Ne isotopes,
although the energies of the highest-occupied neutron s.p.~level
($1d_{3/2}$, $1f_{7/2}$) of $^{27-32}_{\hskip1em K^-}$Ne ($N=17-22$)
decrease due to the additional $K^-$ meson.
This is similar to the case of O isotopes as displayed in Fig.~\ref{f:o}.
In addition, in these larger nuclei,
the deformation changes due to the kaon are much smaller
than those of Be isotopes in Fig.~\ref{f:be}(d).

As mentioned in Sec.~\ref{s:form},
the results of $K^-$ nuclei shown in the previous figures
are obtained with the \kni\ strength $a_0=500 \mfm$.
Because of the uncertainty of the \kni\ strength,
we explore the energies of the highest-occupied neutron s.p.~levels of
$^{27-32}_{\hskip1em K^-}$Ne ($N=17$--$22$)
with \kni\ strengths $a_0=100,200,\ldots,600\mfm$ in Fig.~\ref{f:en},  respectively.
It is interesting to note that the $K^-$ meson does not always increase
the energies of the highest-occupied neutron s.p.~levels
when increasing the \kni\ strength.
In particular,
for the weakly bound $1f_{7/2}$ levels,
a weak \kni\ ($a_0=100,200\mfm$) causes an increase of binding,
whereas repulsion only sets in for $a_0\gtrsim300\mfm$.
This will be analyzed later.

\begin{figure}[t]
\vspace{-2mm}
\centerline{\includegraphics[width=86mm]{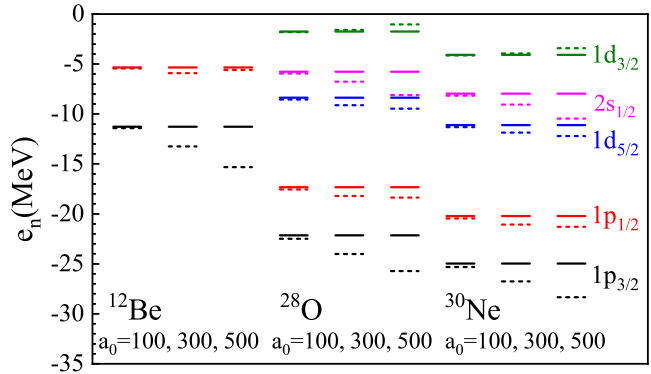}}
\vspace{-2mm}
\caption{
Partial neutron s.p.~levels of spherical nuclei
$^{12}$Be, $^{28}$O, $^{30}$Ne (solid lines)
and their corresponding $K^-$~nuclei (dashed lines)
with $a_0=100$, $300$, and $500\mfm$.
}
\label{f:spb}
\end{figure}

To further illustrate the above feature,
we show in Fig.~\ref{f:spb}
the effect of an additional $K^-$ meson on the neutron s.p.~levels
of the spherical drip-line nuclei $^8$Be, $^{28}$O, and $^{30}$Ne
with three different \kni\ strengths $a_0=100, 300, 500\mfm$, respectively.
Evidently,
the spin-orbit splitting of the orbitals $1p_{1/2,3/2}$ and $1d_{3/2,5/2}$
in $K^-$~nuclei is larger than that in the corresponding normal nuclei.
This effect reduces the binding of the $1d_{3/2}$ valence levels in
$^{28}_{K^-}$O and $^{30}_{K^-}$Ne.
Here we remark that the level inversion between orbitals
$2s_{1/2}$ and $1d_{5/2}$ in light kaonic nuclei obtained by a
RMF model \cite{YANG2019188} is not found in our SHF calculations.

\begin{figure}[t]
\centerline{\includegraphics[width=87mm]{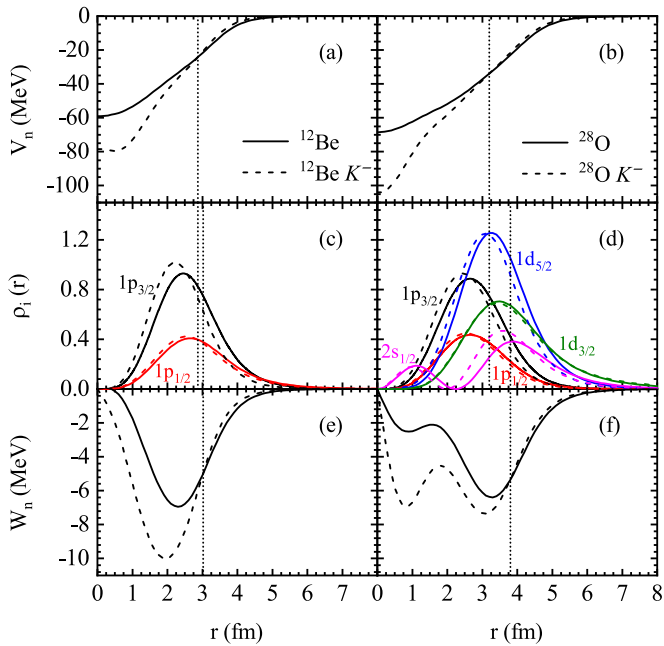}}
\vspace{-3mm}
\caption{
The mean fields $V_n(r)$ and spin-orbit potentials $W_n(r)$,
and the partial densities $\rho_i(r)= 4\pi r^2 v_i^2 |\phi_i(r)|^2$
normalized to the actual occupation numbers of all occupied neutron
s.p.~levels in the normal nuclei $^{12}$Be and $^{28}$O and their
corresponding $K^-$~nuclei with $a_0=500\mfm$.
The vertical dotted lines label the crossings between the $V_n(r)$ and $W_n(r)$
of normal and $K^-$ nuclei.
}
\label{f:rhoi}
\end{figure}

In order to understand the shift of the neutron s.p.~levels by the
addition of a kaon,
we analyze the Schr\"odinger equation~(\ref{e:se}).
One notes that both the central potential $V_n$
and the spin-orbit potential $W_n$ are modified
when including a kaon \cite{zhou2013kaonic,jin2019deformed}.
We remind that the spin-orbit potential of nucleons in the SHF approach is
\cite{PhysRevC.5.626,CHABANAT1998231}
\be
 W_q = \frac{W_0}{2}(\nabla\rho+\nabla\rho_q)
     + \frac{t_1-t_2}{8}J_q - \frac{t_1x_1-t_2x_2}{8}J \:,
\ee
with $q=n$ or $p$.
Qualitatively, the strong $K^-N$ attraction shrinks the nucleus.
This leads to a deepening of the mean fields in the core region of the nucleus,
but a weakening in the peripheral part that is essential
for weakly bound valence neutrons.
Moreover, there is always a delicate competition between $V_n$ and $W_n$
for some levels.
In order to visualize these effects,
we compare in Fig.~\ref{f:rhoi} the potentials $V_n(r)$ and $W_n(r)$
of the drip line nuclei $^8$Be, $^{28}$O, and their corresponding $K^-$~nuclei,
together with the partial densities
$\rho_i = 4\pi r^2 v_i^2 |\phi_i(r)|^2$
of the various occupied neutron s.p.~levels.
One notes that indeed
the attractive \kni\ contracts the density distributions and thus enhances.

We note that $V_n(r)$ and $W_n(r)$ of $^{12}_{K^-}$Be and $^{28}_{K^-}$O
are deeper than their normal nuclei for
$r\lesssim 2.9\,$fm, $r\lesssim 3.0\,$fm, and
$r\lesssim 3.2\,$fm, $r\lesssim 3.8\,$fm, respectively.
The strongly-bound neutron levels
$1p_{1/2,3/2}$, $1d_{5/2}$, and $2s_{1/2}$ of $^{28}$O
are concentrated well within the core region $r\lesssim3\,$fm,
which explains their gain of energy and the larger
splitting of $1p_{1/2}$ and $1p_{3/2}$ in $K^-$~nuclei,
see Fig.~\ref{f:spb}.
Similar phenomena can be found in $^{12}$Be.
On the contrary,
a large amount of the $1d_{3/2,5/2}$-state neutrons locate
in the range of $3.2-3.8\,$fm,
where the central potentials are smaller and the
spin-orbit potentials are larger in $K^-$~nuclei than in the normal nuclei.
Therefore,
the splitting of $1d_{5/2}$ and $1d_{3/2}$ is still enhanced in $K^-$~nuclei.
However, the most peripheral $1d_{3/2}$ neutrons are embedded in weaker
both central and spin-orbit mean fields at $r\gtrsim3.8\,$fm,
which accounts for the upward shift of that level.

In conclusion, the effect of an additional $K^-$ meson on the
neutron drip line depends on the highest-occupied neutron s.p.~level
of nuclei near the neutron drip line.
If this level is an orbit without spin-orbit splitting (e.g., $2s_{1/2}$)
or the lower orbit with splitting (e.g., $1p_{3/2}$ and $1d_{5/2}$),
the additional $K^-$ may extend or not shift the neutron drip line.
If instead this level is an upper orbit with splitting (e.g., $1d_{3/2}$),
the additional $K^-$ can decrease its binding and reduce the neutron drip line,
such as for O isotopes.
This effect is caused by the shrinking of the nucleon wave functions
due to a particularly strongly attractive \kni.
A similar role of an additional $K^-$ in influencing s.p.~levels
was pointed out in the RMF approach \cite{YANG2019188}.
In addition, an extension of the proton drip line is always found
due to added kaon,
in particular because of its negative electric charge.

\section{Summary}
\label{s:end}

We explored the effect of an additional $K^-$ meson on the neutron
and proton drip lines of Be, O, and Ne isotopes
using SHF approach with a simple $K^-N$ Skyrme-type force
and the nuclear SLy4 force including a pairing contribution.
The single-particle levels, binding energies, and quadrupole deformations
were obtained by solving the SHF equations self-consistently.
Due to the attractive \kni,
the additional $1s$-state $K^-$ meson shrinks the nucleon
density distribution and increases its gradient.
This increases the binding of strongly bound inner single-particle orbits,
but might decrease the one of peripheral valence orbits,
also by an enhanced spin-orbit splitting in the kaonic nuclei.
Since the highest-occupied nucleon single-particle levels
and one-nucleon separation energies determine the position of the drip line,
corresponding effects were observed and analyzed in this paper.

We also remark at this point that the treatment of kaonic nuclei
in the static SHF approach can only be approximate,
as their lifetime is very short and an improved dynamical approach
would be required for a more realistic description,
including at least the imaginary parts of kaon wave function and mean field.
In particular, very recently the J-PARC collaboration group
measured the inclusive missing-mass spectrum of the
$^{12}\text{C}(K^-,p)$ reaction and extracted
the potential depth of both real and imaginary parts \cite{Ichikawa2020PTEP}.
Thus, in the future our model can be improved
by devising a more realistic kaon-nucleon Skyrme force
and adding the imaginary part of the kaon optical potential,
aided by current \cite{Ichikawa2020PTEP} and future
experimental data concerning $K^-$~nuclei.
Furthermore, the mean-field approximation employed here
might be inadequate in particular for light nuclei and weakly bound states,
and a beyond-mean-field treatment \cite{Cui17,Cui18,Mei18}
might be required for a more realistic modeling.

\section*{Acknowledgments}

This work was supported by the National Natural Science Foundation
of China under Grants No.~11775081 and No.~11875134, and the
Natural Science Foundation of Shanghai under Grant No.~17ZR1408900.


%

\end{document}